# Room Temperature Dy Spin-Flop Switching in Strained DyFeO$_3$ Thin Films

*Banani Biswas, Federico Stramaglia, Ekatarina V. Pomjakushina , Thomas Lippert, Carlos A.F. Vaz, and Christof W. Schneider\**

B. Biswas, F. Stramaglia, E. V. Pomjakushina, T. Lippert, C.A.F. Vaz, and C.W. Schneider
Paul Scherrer Institute, 5232 Villigen PSI, Switzerland
E-mail: christof.schneider@psi.ch

T. Lippert
Department of Chemistry and Applied Biosciences, Laboratory of Inorganic Chemistry, ETH Zurich, Switzerland

Epitaxial strain in thin films can yield surprising magnetic and electronic properties not accessible in bulk. One materials system destined to be explored in this direction are orthoferrites with two intertwined spin systems where strain is predicted to have a significant impact on magnetic and polar properties by modifying the strength of the rare earth-Fe interaction. Here we report the impact of epitaxial strain is reported on the linear magneto-electric DyFeO$_3$, a canted bulk antiferromagnet with a high Néel temperature (645 K) exhibiting a Dy-induced spin reorientation transition at ≈50 K and antiferromagnetic ordering of the Dy spins at 4 K. An increase in the spin transition of > 20 K is found and a strictly linear, abnormal temperature magnetic response under an applied magnetic field between 100 and 400 K for [010]-oriented DyFeO$_3$ thin films with an in-plane compressive strain between 2% and 3.5%. At room temperature and above, we found that application of ≈0.06 T causes a spin-flop of the Dy spins coupled to the antiferromagnetic Fe spin lattice, whereby the Dy spins change from an antiferromagnetic alignment to ferromagnetic. The spin-flop field gives a lower energy bound on the Dy-Fe exchange interaction of ≈15 μeV.

## 1. Introduction

The control of the electronic properties of materials through growth-induced strain is a well-studied and proven approach to engineer physical properties of materials.[1] In thin films, strain can be introduced in two ways: via chemical pressure, in which an atom is replaced by another atom with different size,[2] or by epitaxial strain, as a consequence of the lattice mismatch between substrate and film.[3] Because the lattice parameters influence the bond length and angle by changing the relative positions of the atoms in the crystalline unit cell, the electronic distribution and their interactions change. Hence, epitaxial strain is a powerful approach to tune the mechanical, chemical, and electrical properties of materials.[4]

Orthoferrites ($R$FeO$_3$) have been investigated extensively since the 1960s for their distinct complex magnetic properties, such as high Néel temperatures (up to $T_N \approx 645$ K for DyFeO$_3$), canted antiferromagnetic (AFM) G-type spin structures with spin reorientation (SR) transitions whose transition temperatures depend strongly on the ionic size of the rare-earth ($R$) element, and ordering of the rare-earth ions at low temperatures.[5] Some of the orthoferrites also show ferroelectric properties.[6] In fact, the coexisting and interacting Fe and $R$ antiferromagnetic spin lattices are expected to lead to polar properties for all $R$ elements except La and Lu, which have zero spin and angular momenta. Recently, this class of materials has gained attention because of their low-temperature magnetic field properties[7-9] and the predicted uniaxial stress-induced ferroelectricity well above room temperature with a large electrical polarization.[10, 11] Such a room temperature multiferroic state would have potential for applications in spin-based electronic devices.

DyFeO$_3$ is one member of the $R$FeO$_3$ family known to have magnetic ordering-driven multiferroic properties below 4 K.[8, 12, 13] It has a distorted perovskite crystal structure (ABO$_3$) characterized by a three-dimensional array of corner-sharing cation-oxygen octahedra of DyO$_6$ and FeO$_6$, with space group Pbnm and tilted Fe-O octahedra along the [001] direction (Figure 1a). The interaction between the two neighboring cations causes three different exchange interactions, namely, Fe$^{3+}$-Fe$^{3+}$, Dy$^{3+}$-Fe$^{3+}$, and Dy$^{3+}$-Dy$^{3+}$, each of which dominates at different temperatures. The strong Fe$^{3+}$-Fe$^{3+}$ interaction causes the ordering of the Fe lattice at 645 K with a predominating G-type AFM ordering along the $a$-axis.[14] However, the G-type ordering is modified by the presence of a Dzyaloshinskii-Moriya (DM) interaction, which leads to a tilting of the oxygen octahedra and a weak ferromagnetic (wFM) component along [001], with A-type AFM ordering along [010]. This spin state is known as $\Gamma_4$ and represented as $\mathbf{G}_x\mathbf{A}_y\mathbf{F}_z$ in Bertaut's notation (Figure 1b).[15] The interaction between Dy and Fe ($Dy^{3+}$-$Fe^{3+}$) increases with decreasing temperature, which eventually leads to a Morin-type SR in the Fe lattice at around 50 K. As a result of this SR, the G-type AFM spin axis switches from the *a*-direction to the *b*-direction, while the wFM component along the *c*-axis disappears concomitant with a change of the magnetic space group. This spin state is known as $\Gamma_1$ and denoted as $A_x\mathbf{G}_yC_z$ (Figure 1c). With decreasing temperature, the $Dy^{3+}$-$Dy^{3+}$ interaction increases and leads to an antiferromagnetically ordered Dy lattice ($T_{N,Dy}$) in the *ab* plane ($G_x A_y$) at around 4 K (Figure 1d).[8, 12] Below $T_{N,Dy}$ and in the presence of a sufficiently large magnetic field applied along the *c*-axis, the Fe spin switches from the $\Gamma_1$ phase to the more stable $\Gamma_4$ phase. In tandem with the change in symmetry, a displacement of the Dy atoms along [001] in the unit cell takes place.[12] This displacement drives the non-polar centrosymmetric unit cell to a polar state. The spin structure of the Fe sublattice has been studied extensively; in contrast, the spin structure of the Dy sublattice is not yet fully understood. In addition, few studies have reported on the interaction between the two magnetic sublattices [7] and how the unit cell distortion impacts this interaction.[11]

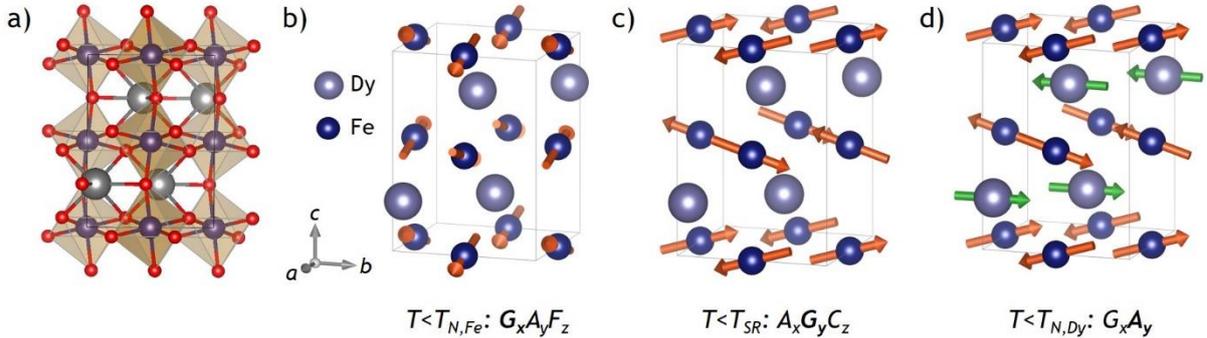

**Figure 1.** a) Crystal structure of $DyFeO_3$. (b) G-type AFM spin structure of the Fe spin-lattice below $T_{N,Fe}$. c) AFM spin-structure of the Fe spin-lattice below $T_{SR,Fe}$. d) AFM spin-structure of the Dy and the Fe spin-lattice below $T_{N,Dy}$.[14, 15]

In this paper, we study the magnetic properties of the Dy spin system in thin $DyFeO_3$ films grown on $YAlO_3$ (010) substrates near room temperature. We show that, while Dy in bulk $DyFeO_3$ is paramagnetic at room temperature,[16] strained thin films exhibit high susceptibility to small magnetic fields applied along the *a*- and *c*-directions. As a result, the spins align either antiparallel or parallel to the field, depending on the field strength, and can be reversibly switched between those two states, resulting in an unusual temperature dependence of the magnetization. The reversible switching behaviour and the independent dual-axis magnetic sensitivity makes strained DyFeO$_3$ an interesting candidate to be explored as a two-axis magnetometer or as a magnetic temperature sensor.

## 2. Results and Discussion

### 2.1. Structural Characterization

The crystalline quality of the deposited films was studied using XRD. All films are [010]-oriented with a full width half maximum of ≈0.02° for the (010) rocking curves. The strain for the [010] out-of-plane direction is tensile and thickness dependent: for thicknesses around 10 nm, it is of the order 2-3% and decreases to < 1% for 100 nm thick films and above. The in-plane strain is compressive for all films and the in-plane axes are aligned with the crystal axes of the substrate. Films with a thickness between 15–20 nm grow coherent with the substrate while thicker films in addition relax towards the bulk DyFeO$_3$ lattice parameter.[17]

### 2.2. Magnetic Properties

At room temperature, DyFeO$_3$ films display a strong magnetic response, as shown in the magnetization loops along [001], [010] and [100] of a 27 nm thin film in Figure 2a. We find a pronounced FM signal and a saturation moment around 8.3 $\mu_B$/f.u. at 5 T. The *M*(*H*) for the [100] direction shows similar results. For comparison, a single crystal exhibits a small FM hysteresis along [001], corresponding to the magnetic soft axis with a saturation moment of 0.06 $\mu_B$ (Figure 2a, inset); no magnetic signal is observed along [100].[9] The origin of the FM hysteresis is the canting of the Fe-O octahedra in the DyFeO$_3$ unit cell along [001].[5] A likely reason for the larger FM hysteresis along [001] and [100] in the strained films relative to a single crystal is the compressive in-plane strain of -3.25 % along the *c*-axis and -2.3 % along the *a*-axis, leading to a stronger tilting and distortion of the Fe-O octahedra; along [010], the strained films are antiferromagnetically ordered and this direction is one of the two hard magnetic axes.[17] The corresponding zero-field cooled (ZFC) measurements (Figure 2b) show, for a field of 0.10 T along [010], a Curie-Weiss (CW) like temperature dependence and a magnetic transition at 6 K. The [100] direction displays a very shallow increase in the magnetic moment between 300 K and 40 K followed by a steep rise and a magnetic transition at ≈6 K. For the [001] direction, a bow-like increase in signal is observed until 70 K, followed by a slow decrease until 10 K and a strong increase down to 3 K.

As shown in Ref. [9], the ZFC measurements for the single crystal along [001] display a spin reorientation transition at $T_{SR}$≈41 K, which can be suppressed easily with increasing field (Figure 3a), in agreement with neutron diffraction measurements.[9] The moment drops by more than 95% within 5 K at $T_{SR}$ and approaches zero with decreasing temperature. Using different probing fields for the ZFC measurements, the moment at $T_{SR}$ changes slightly from 0.15 $\mu_B$/f.u. to 0.16 $\mu_B$/f.u. between 0.05 T and 0.10 T, while $T_{SR}$ is shifted from ≈41 K to below 8 K (Figure 3a). A field of 0.5 T fully suppresses the Fe-spin reorientation by overwhelming the Dy-Fe interaction.[9] For fields >0.5 T, the moment increases considerably and a new magnetic transition, the nature of which is not yet clear, is observed at ≈3 K (Figure 3a, inset).

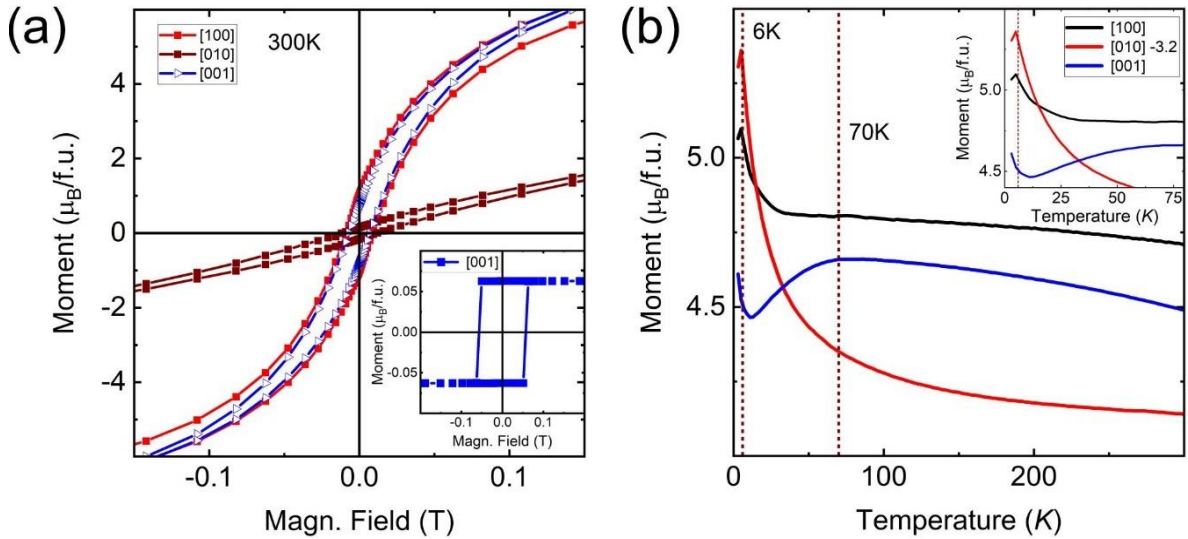

**Figure 2.** a) $M(H)$ for a 27 nm DyFeO$_3$ film at room temperature along [100], [010] and [001]; Inset: $M(H)$ at room temperature for a single crystal along [001]. b) ZFC measurement for [100], [010] and [001] at 0.10 T between 3 K and 300 K. The measurement for [010] has been vertically offset by 3.2 $\mu_B$/f.u. The inset shows a zoomed view of the low-temperature range.

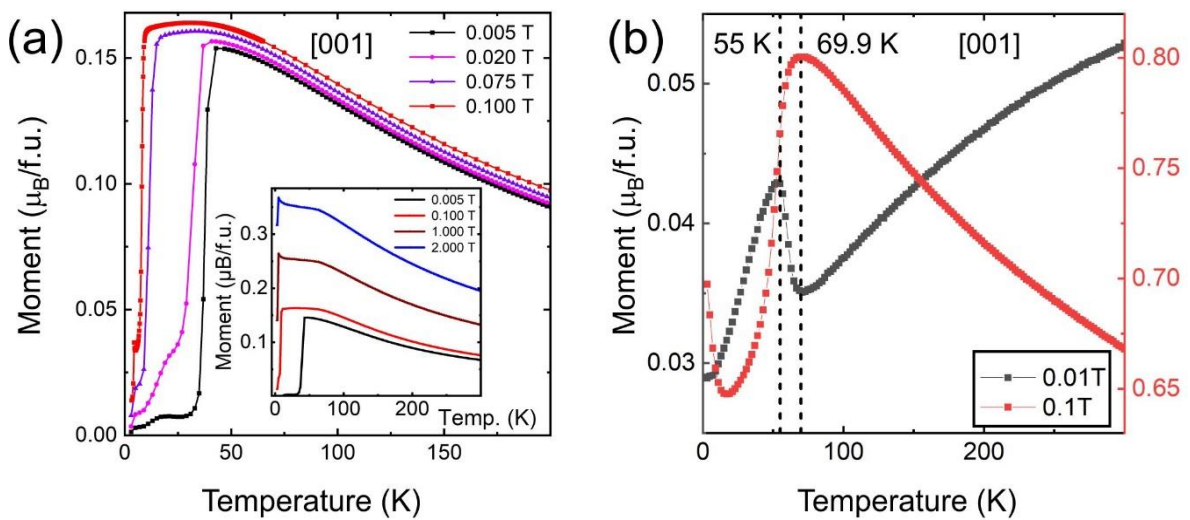

**Figure 3.** a) ZFC measurements along [001] for a single crystal measured at different magnetic fields. Inset shows an extended temperature and magnetic field range. b) ZFC measurements of the 115 nm DyFeO$_3$ thin film along [001] measured with a magnetic field of 0.01 T and 0.10 T. The dashed lines show the transition temperatures at 69.9 K and 55 K.

For films, there are two marked differences relative to the single crystal. First, the sign of the slope depends strongly on the applied magnetic field amplitude for ZFC measurements along [001] and [100] (Figure 3b). The total moment reduces with decreasing temperature, giving rise to a positive slope at an applied field of 0.01 T. A sharp upturn is observed at 69.9 K, followed by a cusp at ≈55 K, with $M(T)$ decreasing down to the lowest temperature measured. In contrast, at 0.10 T, the magnetic moment increases with decreasing temperature until 69.9 K, reduces until 10 K, and increases again at lower temperatures. The second difference to a single crystal is that the transition temperature associated with the spin reorientation increases by 15 K between 0.01 T and 0.0 T for the 115 nm thick film along the [001] direction (Figure 3b). The signal drops by ≈20% over a temperature window of ≈50 K at $T_{SR}$, before rising again with decreasing temperature. The magnetization behavior at higher fields resembles the magnetization curves of the single crystal shown in Figure 3a.

The change in the sign of the slope is observed for the whole film thickness range studied. Figure 4 shows the ZFC measurements for a 51 nm thin film along [100] (Figure 4a) and [001] (Figure 4b) between 100 K and 300 K, where a linear change of $M$ occurs as a function of temperature. We observe a linear increase in the magnetization for both directions as a function of temperature under an applied field of 0.005 T, with a gradient of ≈9×10$^{-6}$ $\mu_B$ K$^{-1}$ along the $a$-direction and ≈6×10$^{-5}$ $\mu_B$ K$^{-1}$ along the $c$-direction. However, a reversal of the slope occurs when the probing field is increased to 0.025 T along [100] and 0.05 T along [001], corresponding to -9×10$^{-5}$ and -8×10$^{-5}$ $\mu_B$ K$^{-1}$, respectively.

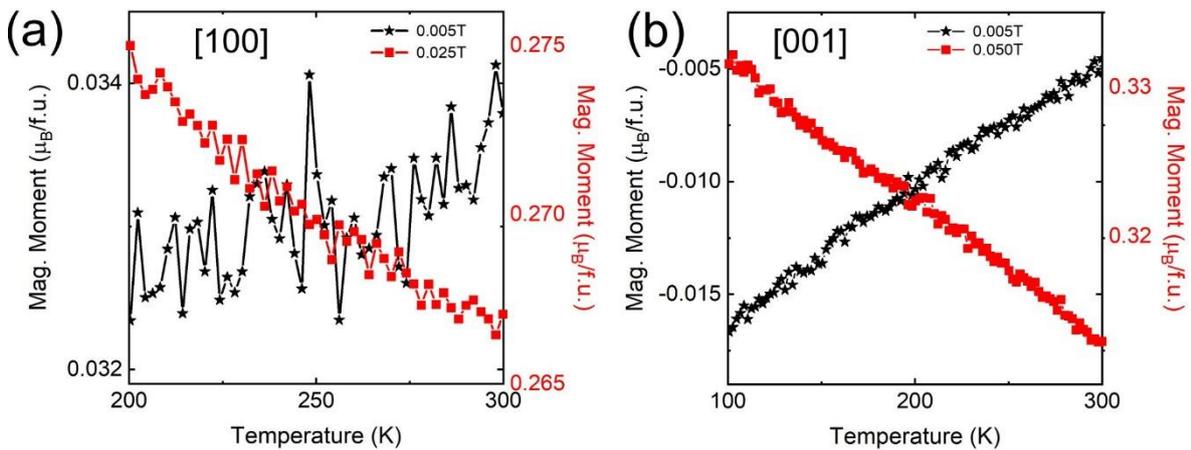

**Figure 4.** Temperature variation of the magnetisation for the 51 nm thick DyFeO$_3$ film measured along a) [100] and b) [001] at different values of the applied magnetic field.

## 2.3. Element-specific Electronic and Magnetic Properties

The change in sign for the $M(T)$ slope of the strained films was unexpected. To gain further insights into the mechanisms responsible for the DyFeO$_3$ magnetic properties, we probed the electronic and magnetic structure of a single crystal as reference and of the 13 nm thick film by means of x-ray absorption spectroscopy (XAS) at the Fe $L_{3,2}$, and Dy $M_{5,4}$ edge at room temperature in total fluorescence yield. For qualitative analysis, spectra taken with left and right circular polarized x-rays were averaged to yield isotropic spectra as a first approximation, while the x-ray magnetic circular dichroism (XMCD) signal is obtained by the difference between the two polarizations. For the magnetic field-dependent measurements, the field was applied along [001] between 0 and 0.10 T.

Comparing the measured Dy $M_{5,4}$ and Fe $L_{3,2}$ spectra qualitatively for the reference single crystal[18] and thin film (Figure 5a), we find little spectroscopic differences for the peak positions and line shapes. The variations in intensities are expected because of differences in self-absorption. The broad agreement indicates a correct valence of +3 for both elements in the film, meaning that the film cation composition is stoichiometric. It is further reasonable to assume that the oxygen content is correct with reference to the DyFeO$_3$ single crystal.[9] The line shapes of Dy and Fe for the film are somewhat different as compared to the single crystal which we attribute to a distorted unit cell, as confirmed from the structural measurements.

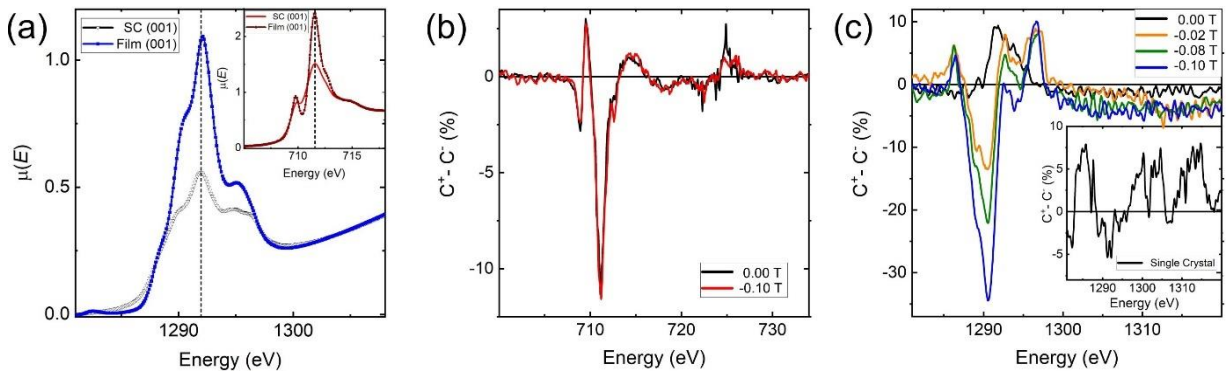

**Figure 5.** a) XAS spectra along the [011]-direction of the Dy $M_{5,4}$ absorption edge for a [010]-oriented DyFeO$_3$ single crystal and a 13 nm thick [010]-oriented film in total fluorescence yield. Inset: the corresponding XAS spectra of the Fe $L_{3,2}$ absorption edge. b) XMCD spectrum for the Fe $L_{3,2}$ absorption edge obtained at 0 T and -0.10 T for a 13 nm thick [010]-oriented film. c) XMCD spectrum for the Dy $M_{5,4}$ absorption edge obtained at different fields for a 13 nm thick [010]-oriented film. The XMCD signal corresponds to the magnetic moment projection along the x-ray beam direction, with the magnetic field applied along [001]. The inset shows the XMCD signal at 0 T for the single crystal.

To determine which cation contributes to the FM signal of the film, a magnetic field is applied parallel to [001], and the x-ray absorption for left and right circular polarized light are measured for Fe $L_{3,2}$ and Dy $M_{5,4}$ edges (Figure 5b, c). The magnetic signature is quite clear and field independent for Fe $L_{3,2}$ (Figure 5b). Likewise, a strong room temperature XMCD signature for the nominally paramagnetic Dy is observed in the thin film (Figure 5c), while the single crystal does not exhibit an XMCD signal (Figure 5c, inset). Since Fe and Dy give a clear and strong XMCD signal, we conclude that both spin lattices contribute to ferromagnetism and to the total moment. In addition, we find that the Dy XMCD signal increases progressively while the Fe XMCD signal remains constant as a function of the applied magnetic field in the range from 0 to -0.10 T. From the $M(H)$ loops in Figure 2a, the moment response is larger along [100] and [001] relative to the [010] direction. The XMCD results indicate that the remanent moment is from Fe and therefore, suggest that the Dy spins respond along the directions where the Fe moment is present. This means that the magnetic response of Dy is induced by the Fe net moment and the energetically favoured spin state is ferromagnetic. This resembles a transition from a speromagnetic state to an asperomagnetic state, but for low fields only a portion of the Dy spins is involved in attaining a FM state.[19, 20]

Field-induced magnetism in rare-earth ions that are paramagnetic in the unstrained state has been reported for GdFeO$_3$ and ErFeO$_3$.[21, 22] The calculated effective fields for GdFeO$_3$ are on the order of 1-2 T, at most 0.10 T are noted for strained films above $T_{N,R}$. Looking into $M(H)$-loops of the single crystal,[9] it becomes clear that the induced magnetism in the paramagnetic Dy also takes place above $T_{N,Dy}$ along [001] with a FM alignment of the Dy with respect to Fe for a switching field of ≈0.05 T at room temperature. The reported double step for films along [100] and [001] associated with the switching of the Dy spin becomes visible near $T_{SR}$ for the single crystal and is otherwise superimposed by the FM hysteresis at higher temperatures.[9] For powders, signatures in the $M(H)$ similar to that of single crystals can be measured for rather small applied magnetic fields.[9] We therefore conclude that the induced magnetism in the paramagnetic rare-earth system is a general feature for orthoferrites as pointed out in ref. [21].

These observations confirm that Dy is magnetically active at room temperature and that the varying component of the $M(H)$ hysteresis loops arises from the Dy moments while the remanent value corresponds to the Fe net moment. The Dy spin flops into the corresponding Fe-spin direction in order to minimize the magnetic exchange energy. Hence, we conclude that the double-step hysteresis $M(H)$ loops along [100] and [001] measured between 1.6 K and 390 K originate from the coupling between the Fe and Dy spins.[17] It is further interesting to note that the change from an AFM to a FM Dy spin arrangement as a function of the applied field is of a gradual nature for small fields rather than a sharp spin transition. This implies that a finite number of Dy spins are involved at a given field until at ≈1 T all Dy spins become FM aligned and the step in the *M(H)*-loops occurs.[17]

**2.4. Discussion**

From the above bulk and element specific magnetometry results, the following picture of the magnetic behaviour of the strained DyFeO$_3$ films emerges.

2.4.1 Dy Spin Flip-Flop

i) We start by considering the fact that, in the single crystal, the Fe-spins point along [100] at higher temperatures and the spin direction changes to [010] at $T_{SR}$, while Dy orders in the *ab*-plane at $T_{N,Dy}$. The spin state of the Fe lattice above $T_{SR}$ is in the $\Gamma_4$ phase, characterized by a weak FM ordering along [001]. A change in magnetic symmetry occurs at the spin reorientation transition from $\Gamma_4$ to $\Gamma_1$, coupled with a cancelling of the wFM component along the *c*-direction.[5, 6] The behaviour for strained films is strikingly different: a significant ferromagnetic component is observed in strained films along [100] and [001] (see Figure 2a) and the change from $\Gamma_4$ to $\Gamma_1$ does not take place.[17] Above and below $T_{SR}$ the films have a FM component along [001] and [100] which does not vanish down to the lowest measured temperature of 3 K. The Dy ordering is in the *ab*-plane, as expected; however, an additional component along the *c*-direction (Figure 2b) emerges when measuring *M(T)* at larger fields. The Dy ordering temperature is anisotropic and sensitive to small magnetic fields. The increase in $T_{SR}$ in thin films is a consequence of compressive strain and the large transition width is attributed, in part, to the strain anisotropy.[17]

ii) In bulk DyFeO$_3$, the spins of the Fe lattice contribute most to the net moment between $T_{N,Fe}$ and $T_{N,Dy}$, and *M(T)* can be described best by a sum of Curie-Weiss $(T+T_N)^{-z}$ terms for Fe and Dy. For the fitting, $T_{N,Fe}$=650 K and $T_{N,Dy}$=4 K are fixed, while *z* is a free parameter for the Dy-term and is expected to be 1 for Fe. For the *a*-direction between 100 and 300 K, the CW-relation cannot be fitted well to the measurements. The *b*-direction is showing largely the expected CW-relation. Plotting the inverse moment vs *T* provides a clearer picture (Fig. S1). For all three directions, there are linear slopes at higher temperatures with a positive intercept at *T*=0 directly indication a net FM interaction between Fe and Dy.[23]

iii) For strained films, a magnetic field applied along the *c*- or *a*-axis induces a net moment in the Dy spins. Initially, the ordering is antiparallel to the Fe net moment up to a critical magnetic field value, above which it flips to align parallel to the Fe spins, resembling a flip-flop transition. This argument is supported by the Dy XMCD signal and schematically illustrated in Figure 6a. The antiparallel alignment between the Dy and Fe net moment at low field values is responsible for the positive temperature slopes and represents a low energy state.[7] The magnetic state changes to a parallel spin arrangement (high energy state) when the applied magnetic field exceeds a critical field, inverting the *M*(*T*) slope to negative (Figure 6b). We note that $(dM/dT)_H = (dS/dH)_M$ (*S*: entropy), hence, at $(dM/dT)_H = 0$ implies a maximum in the magnetic entropy with the applied magnetic field, which we associate to a magnetic (phase) transition associated to the proposed spin-flop transition (Figure 6b inset). The Dy spin-flop is therefore explained in terms of the interaction between the net Fe spin moment lattice and the Dy spin system. The experimentally determined critical field value for the 13 nm film is ≈0.06 T (see Figure 6c) and is similar for the 115 nm films, suggesting that this phenomenon does not depend strongly on film thickness or partial strain relaxation.

2.4.2 Dy Spin-Flop Properties

The origin of the magnetically active Dy spins well above $T_{SR}$ is most likely related to strain and possibly a result of the displacement of Dy within the $DyFeO_3$ unit cell; for the 13 nm film the strain components for the three crystallographic orientations correspond to -2.3%, 2.5% and -3.46%, whereby 1% lattice strain is equivalent to a pressure of 1 GPa. In compounds structurally similar to $DyFeO_3$, the Dy ordering temperature is typically in the range of a few K, for example, for $DyMnO_3$.[24] The introduction of a compressive strain component along [100] and [001] and tensile strain along [010] results in a magnetic anisotropy not present in single crystals, which is manifested as a step in the *M*(*H*) loops along [100] and [001].[17] Another consequence of the magnetic anisotropy is the Dy ordering temperature increase to 5 and 6 K for the *a*- and *b*-direction, and lower for the *c*-direction. At present, we find no clear systematic behaviour for the Dy ordering temperature with film thickness, but in general $T_{N,Dy}$ is larger than reported for single crystals.

In the Fe-Dy two-spin system, Dy is paramagnetic below $T_{N,Fe}$ and the magnetic state is formed from randomly placed magnetic moments locked in various orientations below an ordering temperature. We have already shown that the magnetic interaction energy is positive and the axial crystal field strength for an f-shell ion will be at least as large or bigger than the magnetic interaction energy.[19] If the crystal field strength is larger than the exchange energy, the magnetic state is less rigid and more susceptible to external fields and the magnetisation is anisotropy-dominated. Under an applied magnetic field, the energetically favorable alignment of the moments becomes ferromagnetic and presents a lower energy state where the spins flip into.[19, 20]

Stable and reproducible switching of the Dy spin at room temperature and above would enable strained $DyFeO_3$ to be used as a magnetic field sensor. A technically interesting point is that the selected film geometry presents a natural two-axis magnetometer. This is shown in Figure 6b when measuring $M(T)$ with different bias fields along [001]. Here, the magnetization data for the different bias fields has been adjusted by adding a constant offset to make the different field-dependent measurements more comparable for a qualitative analysis, using the room temperature value of 5.21 $\mu_B$/f.u. measured for the 13 nm film at 0.10 T as the common reference point. For small fields, between 0.005 T and 0.025 T, $M(T)$ decreases almost linearly from room temperature to 100 K, where a stronger decrease takes place due to the appearance of the spin reorientation arising from an increase of the $Fe^{3+}$-$Dy^{3+}$ interaction. For a field of 0.075 T and 0.10 T, the slope for $M(T)$ rises initially with decreasing temperature and starts to drop rapidly at around 100 K.

The magnetic sensitivity of Dy to small magnetic fields and the linear $M(T)$ dependence over such a large temperature range are the consequence of a mutual compensation of Fe and Dy moments. Depending on the bias field, the temperature window between 100 and 400 K has a linear relationship between $M$ and $T$ with a gradient of 0.02 T K$^{-1}$. In addition, there is a large magnetic susceptibility, particularly evident at room temperature, which is puzzling. With an applied field of 0.10 T, non-interacting paramagnetic Dy ions will not result in such large magnetic moments over the measured temperature range. Furthermore, the exchange between $Fe^{3+}$ ions in a $d^5$ electronic configuration can only be antiferromagnetic and the relatively weak spin canting due to DM interactions is not known to yield large moments. We can further exclude a large exchange field on Dy ions at room temperature and above, since the Fe-$R$ exchange interactions for a single crystal are not strong (≈10 µeV) and Fe spins have a tendency to order antiferromagnetically.[17, 25] One experimental observation is a finite conductivity at room temperature giving rise to substantial leakage currents when probing ferroelectric properties. Conducting electrons, however, mediate double exchange interactions between spins, which can result in large magnetic moments. Alternatively, an orbital magnetization contributes to the total moment. One reason for the presence of some conduction electrons can be a charge transfer from $O^{2p}$ to $Fe^{3d}$ orbitals.[26] A relatively low room temperature resistance for orthoferrite single crystals with a band gap of 2.1-2.3 eV is known and likely to be reflected in electrical film properties.[18] In this case, $DyFeO_3$ at room temperature behaves electrically like a lightly doped semiconductor. Alternatively, the strong anisotropic distortions of the film lattice and enhanced exchange interactions could be the origin of this unusual magnetic behaviour.

2.4.3 Dy Spin Flop Sensor

As shown in Figure 4, a temperature gradient of $10^{-5} \mu_B$ K$^{-1}$ gives an indication of the sensitivity of such a device. Plotting the linear slope of $M(T)$ vs. applied field shows directly the Dy spin behaviour with field (Figure 6b, inset). Already at 0.025 T, the Dy spins start to turn away from an antiparallel alignment with increasing field before crossing the zero-line at 0.06 T for both crystalline directions. The changeover between a falling and a rising slope above room temperature also takes place for a magnetic field of ≈0.06 T (Figure 6c). A magnetic field balancing the influence of the Fe and Dy spins on $M(T)$ gives a lower limit on the Dy-Fe exchange energy in thin films. For a field of 0.06 T, this corresponds to ≈15 µeV and this is on the same order of magnitude as the Dy-Fe exchange energy of ≈10 µeV for a $DyFeO_3$ single crystal, but much smaller than the ≈100 µeV required to overcome the Dy-Dy interaction.[9] Reproducible switching between the low and high energy spin states is shown in Figure 6d when measuring $M(T)$ at 0.01 T (black symbols) and 0.1 T (red symbols) between 300 and 350 K repeatedly. For the 0.01 T measurement, the slope with increasing temperature is positive, whereas for the larger field, the slope is negative.

With a gradient of 0.02 T K$^{-1}$ the magnetic field sensitivity is expected to be equivalent to Hall sensors and hence strained $DyFeO_3$ films can be used as an on-chip field sensor. This is a thin film alternative to mixed Tb/Ho orthoferrites where the onset of the SR is a function of the Tb/Ho ratio.[27] Whether the sensitivity can be similar to flux gate sensors will depend on the detection and read-out scheme.[28] A direct switching with a small reference field would yield a dynamic change of the magnetic moment of at least 10-fold if a change in the spin direction is measured. If field-biased at the bistable field, a significantly larger sensitivity can be achieved using a differential measuring scheme. Both crystalline directions show the same zero-crossing for $\Delta M/\Delta T$ vs. applied field. This would also allow for a simultaneous two-axis gradiometer read-out scheme.

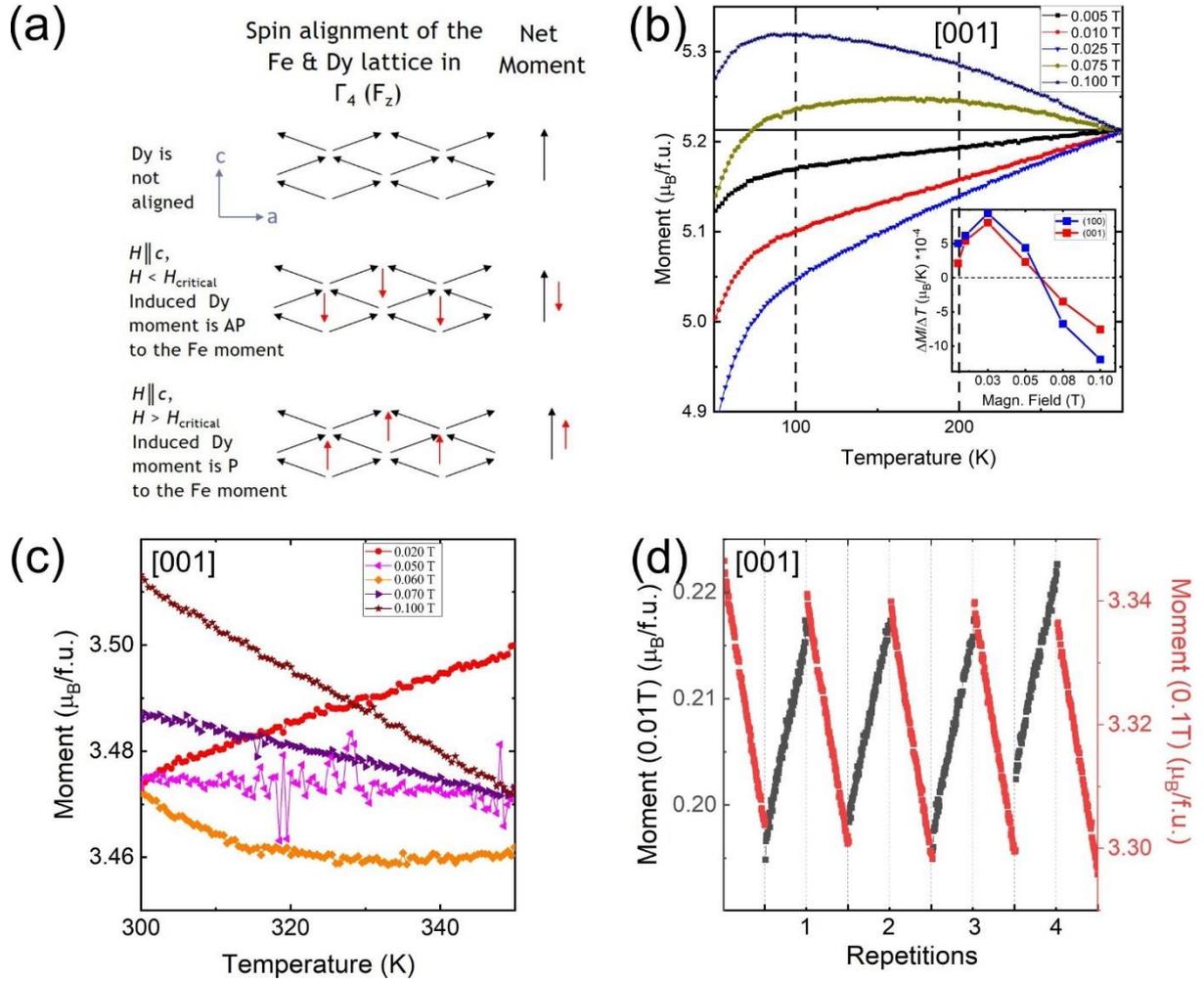

**Figure 6.** a) Model showing the spin alignment of the Fe and Dy lattice in $\Gamma_4$ ($F_z$) for strained films with the magnetic field. b) ZFC $M(T)$ between 50 K and 300 K for different applied magnetic fields for the 13 nm $DyFeO_3$ film. The ZFC measurements have been shifted with respect to the room temperature moment of the 0.10 T measurement (0.005 T: 5.09 $\mu_B$ 0.01 T: 4.59 $\mu_B$, 0.025 T: 3.08 $\mu_B$, 0.075 T: 0.48 $\mu_B$). The inset shows the variation of $\Delta M/\Delta T$ over the temperature range indicated by the vertical dashed lines. c) Determination of the magnetic field required to reach $M(T)$=constant for the 13 nm $DyFeO_3$ film above room temperature. d) $M(T)$ between 300 and 350 K measured repeatedly with a field of 0.10 T (red) and 0.01 T (black) along [001] for the 13 nm $DyFeO_3$ film.

In the investigated orthoferrite system, Dy is the element with the largest magnetic moment, becoming magnetically active at room temperature when deposited as a strained thin film. As already indicated and theoretically described,[21] we anticipate that Dy can be replaced with other rare-earth elements in orthoferrites except Lu and La, to give comparable magnetic responses when prepared as thin films. Since the growth-induced strain will affect the structural and magnetic symmetry of the unit cell, these strained orthoferrite films could exhibit ferroelectricity due to a lowering of the crystal symmetry, resulting in a multiferroic state. The coupling strength between magnetic and ferroelectric order parameters is expected to be large, since strained DyFeO$_3$ is a type 2 multiferroic material, i.e., where the ferroelectric order is driven by the spin state. The magneto-electric coupling in such a system would then enable electric read-out when observing a change in the magnetic response.

## 3. Conclusion

In summary, we have shown here the presence of an abnormal magnetic response in strained [010]-oriented DyFeO$_3$ thin films grown on YAlO$_3$(010) substrates, with a compressive in-plane strain between 2% and 3.5%, resulting in a 2% tensile strain out-of-plane. Applying small magnetic fields along the [100] and [001] direction causes a spin-flop of the Dy spin coupled to the antiferromagnetic Fe spin-lattice at room temperature and above, extending down in temperature down to the Dy ordering. The presence of magnetic activity in Dy was shown by measuring the XMCD signature at the Dy M$_{5,4}$ edge. The origin of the enhanced magnetic activity of Dy is not yet clear, but is attributed to the highly strained unit cell with modified Fe-Dy interaction and enhanced magnetic anisotropy. As a consequence, Dy shows a spin-flop along the [100] and [001] direction but not along [010]. The Dy spin reversal also gives a lower limit to the Dy-Fe exchange interaction of ≈15 µeV. We have further shown, that coherent b-axis oriented DyFeO$_3$ thin films are natural two-axis gradiometers with a potential electric read-out when sensing changes to a magnetic response due to the multiferroic nature of this class of materials.

## 4. Experimental Section/Methods

Epitaxial thin films of DyFeO$_3$, with bulk lattice parameters $a$=5.302 Å, $b$=5.598Å, and $c$=7.632Å, were grown by pulsed laser deposition using a KrF excimer laser ($\lambda$ = 248 nm, 3 Hz) with an in-plane lattice mismatch between DyFeO$_3$ (010) and a YAlO$_3$ (010) substrate ($a$=5.180 Å, $b$=5.330 Å, and $c$=7.375 Å) of -2.35 % (compressive) along the $a$- and -3.25 % (compressive) along the $c$-axis.[29] The laser beam profile, defined by an aperture, is focused onto a sintered ceramic target with a spot size of 1.4 × 1.4 mm$^2$ and a laser fluence of 2.0 J cm$^{-2}$. The deposition was performed in an O$_2$ background at 0.33 mbar, a substrate temperature of 700°C and a substrate-target distance of 5 cm.[29]

The structural quality of the films and their thickness after the deposition was measured by x-ray diffraction (XRD) and x-ray reflectometry (XRR), respectively, using a Seifert 3003 PTS diffractometer with a Cu $K_{a1}$ monochromatic x-ray source. θ−2θ scans were used to monitor the crystalline quality of all samples, reciprocal space maps to determine in-plane lattice constants and to evaluate qualitatively the strain in these films, and XRR to determine the film thickness after the deposition. Details of the structural characterisation have been reported in Ref. [17] and the lattice constants for the different films, bulk $DyFeO_3$ and $YAlO_3$ are provided in the SI. Prior to the deposition of the thin film series, the composition was verified using Rutherford backscattering (RBS) to ensure that the selected deposition yields a Fe/Dy ratio of 1±0.02. The oxygen signal from the film could not be separated reliably from the substrate contribution.

The temperature dependent magnetization was investigated by a commercial SQUID magnetometer (Quantum Design, MPMS® 3). The evolution of the magnetic moment ($M$) with temperature were analysed using the following measurement protocol. First, the sample is cooled down without applied field to 3 K. The zero-field cooling (ZFC) $M(T)$ measurement was taken during heating between 3 K and 300 K in the presence of a small field applied along the measurement direction. After each measurement cycle, the film was kept at 390 K for 5 min., which is close to the maximum temperature available; while ideally the temperature should be higher than the magnetic critical temperature (~ 645 K for bulk $DyFeO_3$ but below 600 K for strained films[8]), the increased thermal fluctuations are expected to fully demagnetise the sample. This procedure provided reproducible nominal ZFC $M(T)$ measurements even after measuring $M(H)$ up to 7 T. After this thermal protocol, the samples were cooled down without an applied field and the measurements were taken during warming (ZFC).

X-ray absorption (XAS) measurements of the single crystal and of the 13 nm thick film were performed at the SIM beamline of the Swiss Light Source.[30] The Fe $L_{3,2}$-edge, the O $K$-edge as well as the Dy $M_{5,4}$ XAS data were acquired at room temperature with the beam incident at 45° to the sample surface for left and right circularly polarised light. The measurements were done in total fluorescence yield (TFY).[31] Prior to the measurements, the surface of the crystal was cleaved to collect spectra from a fresh surface for reference purposes.[18] All spectra were normalised to the photon flux recorded by a clean gold mesh upstream of the experimental chamber. The corrected spectra were normalized using the background edge jump of the regions well below and beyond the $L_{3,2}$ (for Fe), and $M_{5,4}$ (Dy) absorption edges. All XAS data treatment was done using the Athena and Origin software.[32]

**Acknowledgements**

This work was supported by the Swiss National Science Foundation (Project No. 200020_169393 and 200021_184684) and the Paul Scherrer Institute. Part of this work was performed at the Surface/Interface: Microscopy (SIM) beamline of the Swiss Light Source, PSI. The authors would like to thank V. Michel for helping to prepare Figure 1 and Nikita Shepelin for critical reading the manuscript. The authors further would like to thank M. Mostovoy, M. Fiebig and E. Bousquet for discussions on the magnetism of Fe and rare-earth elements in orthoferrites.

# Supplementary Information

**Room Temperature Dy Spin-Flop Switching in Strained DyFeO$_3$ Thin Films**

*Banani Biswas, Federico Stramaglia, Ekatarina V. Pomjakushina, Thomas Lippert, Carlos A.F. Vaz, and Christof W. Schneider\**

B. Biswas, F. Stramaglia, E. V. Pomjakushina, T. Lippert, C.A.F. Vaz, and C.W. Schneider
Paul Scherrer Institute, 5232 Villigen PSI, Switzerland
E-mail: christof.schneider@psi.ch

T. Lippert
Department of Chemistry and Applied Biosciences, Laboratory of Inorganic Chemistry, ETH Zurich, Switzerland

S1

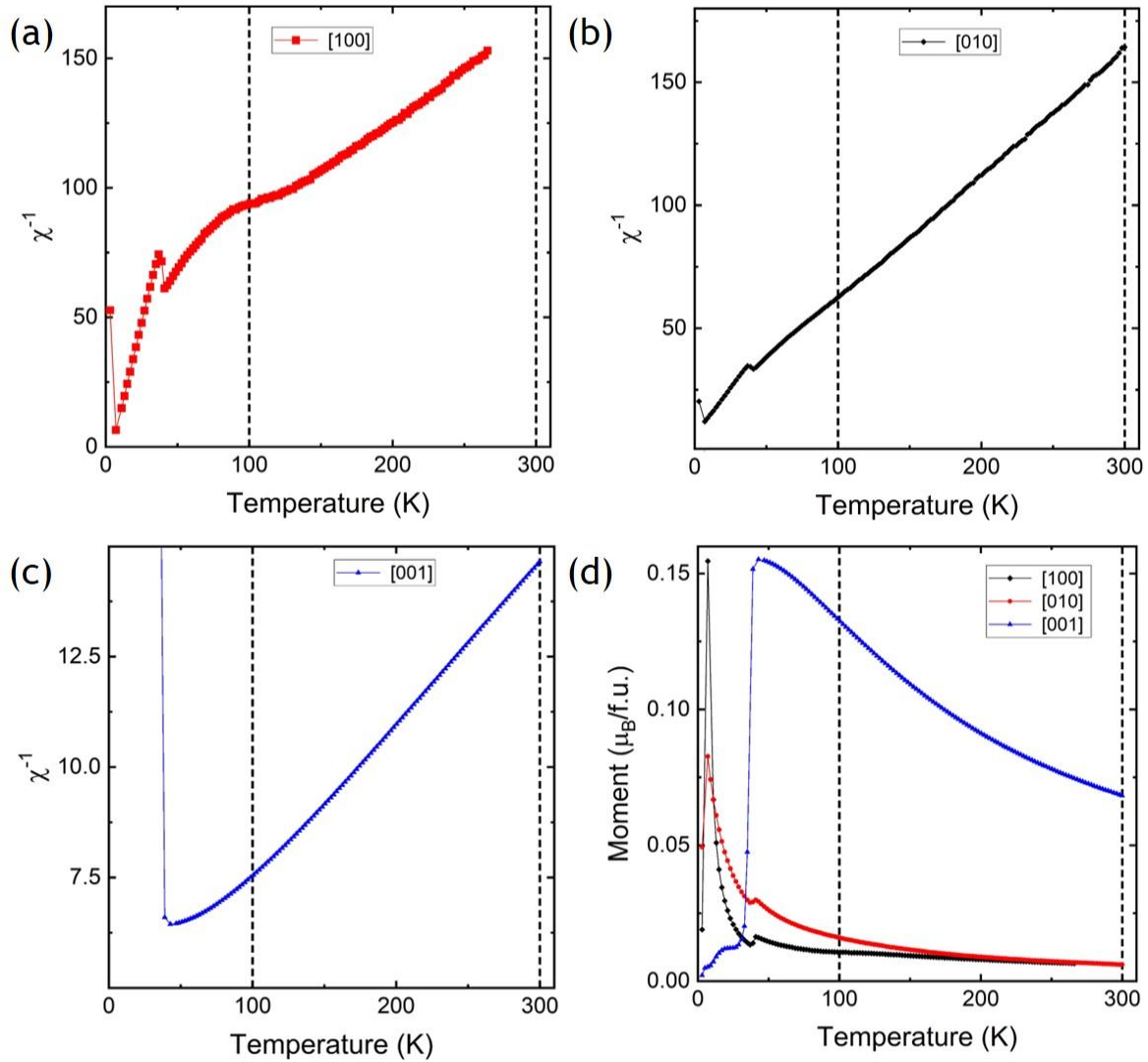

Figure S1: Inverse moment vs temperature for the DyFeO$_3$ single crystal for (a) the [100], (b) the [010] and (c) the [001] direction. (d) all three $M(T)$ measurements in one graph as an overview of the full temperature range measured. The temperature interval between 100 K and 300 K is the window where the influence of the induced Dy moment is expected to be small compared to the Fe moment.

Lattice constants for (010)-oriented DyFeO$_3$ films grown on (010) YAlO$_3$

| Thickness (nm) | [100] in Å | [010] in Å | [001] in Å | uc-volume in Å$^3$ |
| --- | --- | --- | --- | --- |
| 13 | 5.182 | 5.748 | 7.373 | 219.622 |
| 27 | 5.184 | 5.689 | 7.447 | 219.625 |
| 51 | 5.276 | 5.645 | 7.675 | 228.585 |
| 78 | 5.311 | 5.631 | 7.655 | 228.932 |
| 115 | 5.32 | 5.631 | 7.622 | 228.331 |
| | | | | |
| DyFeO$_3$ | 5.302 | 5.598 | 7.632 | 226.255 |
| YAlO$_3$ | 5.18 | 5.33 | 7.375 | 203.619 |